# Balancing the Byline:

# Exploring Gender and Authorship Patterns in Canadian Science Publishing Journals

Eden J. Hennessey[1], Amanda Desnoyers[1], Margaret Christ[1], Adrianna Tassone[1], Skye Hennessey[1], Bianca Dreyer [1], Alex Jay[1], Patricia Sanchez[2], and Shohini Ghose[1, 3]

[1] Wilfrid Laurier University, Centre for Women in Science (WinS)

[2] University of Toronto, Department of Psychology

[3] Quantum Algorithms Institute

Corresponding author:

Eden Hennessey (ehennessey@wlu.ca)

## Abstract

Canada is internationally recognized for its leadership in science and its commitment to equity, diversity, and inclusion (EDI) in STEM (science, technology, engineering, and math) fields. Despite this leadership, limited research has examined gender disparities in scientific publishing within the Canadian context. This study analyzed over 67,000 articles submitted to 24 Canadian Science Publishing (CSP) journals between 2010 and 2021 to better understand patterns of gender representation. Findings showed that women accounted for less than one-third of published authors across CSP journals. Representation varied by discipline, with higher percentages of women in biomedical sciences and lower percentages of women in engineering—trends that mirror national and global patterns. Notably, the percentage of women submitting manuscripts closely matched those published, suggesting that broader workforce disparities may play a larger role than publication bias. Women were less likely to be solo authors or hold prominent authorship positions, such as first or last author—roles typically associated with research leadership and career advancement. These findings point to the need for a two-fold response: continued efforts to address systemic barriers to women's participation in STEM, and a review of publishing practices to identify effective strategies that ensure equitable access, recognition, and inclusion for all researchers.

## Introduction

Canada has long been considered a global leader in science, technology, engineering, and mathematics (STEM) research, contributing significantly to advancements in fields such as quantum computing, artificial intelligence, environmental science, and medical innovation (Chief Science Advisor Annual Report, 2022-23; Innovation, Science and Economic Development Canada, 2022). According to Nature Index data, Canada ranked 7th of the top 50 countries named research leaders in 2023 (Nature Index, 2024). Alongside the country's scientific excellence, Canada is also perceived as a global leader in advancing equity, diversity, and inclusion (EDI) in STEM, notably through commitments such as the Dimensions Charter (Government of Canada, 2023) and the federal focus on EDI within Canada's research funding landscape (e.g., Canada Research Chairs Program, CRC, 2023; Canada Research Chair EDI action plan, Government of Canada, 2025a; NSERC Chairs for Inclusion in Science and Engineering, NSERC, 2025a; NSERC Chairs for Women in Science and Engineering, NSERC, 2025b; specific funding for EDI initiatives in academia such as the Canada Research Chairs EDI stipend, Government of Canada, 2025b). These initiatives require leadership, funding, and staff to implement and therefore represent a substantial commitment of financial and human resources. While Canada makes significant investments in supporting inclusion in scientific research, it is not immune to the global trend of gender disparities in scientific publication. However, very little research has assessed possible gender bias in STEM publishing in the Canadian context specifically.

Reducing bias and promoting access in science is not only a social justice imperative but can strengthen research by fostering a range of perspectives, ideas, and approaches, leading to more innovative and impactful discoveries. Some research has directly linked gender diversity in research teams to increased scientific productivity. Yang and colleagues (2022) studied mixed-gender research teams, examining 6.6 million papers published across the medical sciences since 2000. Results showed that the publications of mixed-gender teams were substantially more novel and impactful than the publications of same-gender teams of equivalent size, and the more gender balance on teams, the higher the scores on performance outcomes. By promoting environments where researchers, inclusive of all dimensions of diversity (e.g., gender, race, etc.) can thrive, STEM institutions—including publishers—can foster a research ecosystem that produces high-quality and innovative research (Campbell et al., 2013; Hofstra et al., 2020; Yang et al., 2022). Given the critical role of publishing for scientific advancement and scholarly recognition, assessing demographic representation and inclusion in academic publishing has become a focus for researchers and policymakers alike (Else & Perkel, 2022).

Previous studies have identified various levels of gender disparities in STEM authorship across disciplines. Larivière et al., (2013) analyzed over five million papers across various disciplines and found that women authors are significantly underrepresented in fields like physics, computer science, and mathematics. Disciplines like health sciences and social sciences have a higher representation of women authors, though a publication gap still exists. West and colleagues (2013) conducted an analysis of gender differences in publishing across a wide array of disciplines, revealing that women were particularly underrepresented in physical sciences and engineering, whereas they fared better in life sciences and social sciences. More recently, Jaramillo and colleagues (2025) analyzed 80 million papers published from 1975 to 2020 in 19 academic fields. They found that women were underrepresented as authors in all fields. The gender gap was particularly pronounced in physics, geology, and mathematics, and was lower –but still present—in social sciences like psychology. This gender gap in scientific publishing mirrors underrepresentation trends in particular STEM fields (Huang et al., 2020; Holman et al., 2018; Paule-Vianez et al., 2020). In general, analyses of gender bibliometrics in scientific research underscore that the disparity in authorship is wider in fields like physics, engineering, and computer science, while life sciences, social sciences, and health-related fields show smaller, though still present, disparities.

Disparities between men and women also appear in team composition and authorship order (i.e., order in which author names appear on a paper) across scientific disciplines. For example, the gender gap in solo versus group authorship in STEM has been the focus of several studies, revealing inequities in the representation and recognition of women in collaborative and solo-authored works. Notably, previous research shows that women are significantly underrepresented as solo authors, particularly in fields like physics and engineering (Larivière, et al., 2013; West et al., 2013; Rorstad & Aksnes, 2015; Mauleón et al., 2013; Day et al., 2020). In group-authored papers, women are more likely to occupy middle-author positions, suggesting less visibility in leadership roles, which typically correspond to first or last authorship. Hallas (2025) assessed last author gender in submissions to Nature Physics over the last decade and found across 1,804 original research reports published over ten years, only 142 (8%) had a woman last author while 1,575 had a man last author (87%). Studies have also demonstrated that group collaborations are more common for women, though they often face challenges (e.g., work is unrecognized or ignored, inequities in the division of scientific labour) in being credited as lead or senior authors (Ross et al., 2022; Holman et al., 2018; Macaluso et al., 2016).

It is also noteworthy that the order in which author names appear on an article may have different meanings depending on discipline. For example, in economics, authors are listed alphabetically, whereas in medicine or biology, the first author position implies the largest contribution, and the last author position infers a senior researcher who provided funding and supervision (Hundley et al., 2013), although the meaning of last authorship varies by discipline and journal. Previous large-scale bibliometric analyses have revealed an association between authorship order and contributions (Larivière et al., 2021), such that in science fields, first and last author roles were consistently associated with the largest proportion of contributions, compared to middle authors, whose contributions were more likely to be technical (Larivière et al., 2016; Sauermann & Haeussler, 2017).

Internationally, women are underrepresented in senior authorship roles, prestigious journal publications, and citation metrics, calling attention to a gap in visibility and recognition that affects career progression and funding opportunities (Elsevier, 2017; Holman, et al., 2018). Such gender disparities in authorship are present globally, with women often encountering barriers to publication and recognition that can severely impact academic career advancement (Huang et al., 2020). Publishing is a critical aspect of maintaining leadership in research, as it is the primary means by which scientific careers are built (Lee, 2019; Lindahl, 2018). Women with fewer or lower-impact publications (e.g., lower citations, published in journals with a lower-impact factor, less public engagement) may struggle to secure research funding, which in Canada, women are less likely to apply for or be awarded in the first place (Natural Sciences and Engineering Council of Canada; NSERC, 2024), given that past research performance is used as criteria for grants (e.g., the NSERC Discovery Grant; NSERC, 2025c). In competitive STEM fields, a strong publication record is essential not only for securing grants but also for fostering innovation and translating research into real-world applications. Publications with academic impact (e.g., measured through citation counts, high-impact factor journals, novelty) and social impact (e.g., measured through impact on policy, innovation, public engagement) are critical to sustaining leadership in the rapidly evolving landscape of global science.

Research contributions beyond those in the traditional format of publications are increasingly recognized as valuable. The Declaration on Research Assessment (DORA) considers alternative outputs such as data sets, software, code, and impact on policy to be important research contributions and provides guidance and advocacy on using these alternatives to assess research for hiring, promotion, and funding decisions. DORA has over 25,000 signatories as of May 2024 (Declaration on Research Assessment; nd) and there is evidence for its impact on research assessment within organizations (Saenen et al., 2021). However, one's publication record continues to be a crucial component of academic career progression (Mantai & Marrone, 2023) and therefore an important area of analysis.

Considering Canada's substantive financial and social investments in supporting gender equity and inclusion in science (NSERC 2025a, 2025b; Government of Canada, 2025a, 2025b) it is important to assess and monitor all aspects of the research landscape—including publication—for representation and potential bias and barriers. However, there are two separate issues to consider in addressing this inquiry; first, the extent to which women are represented in various scientific fields around the globe, and second, the practices of publishing firms in promoting gender inclusion and removing potential bias in the publication process. For example, the use of double anonymized reviews, EDI best practices for hiring editors, and providing instructions that mitigate bias to reviewers (Jack et al., 2023). If the percentage of women authors in submitted and published articles is similar, then women's underrepresentation in scientific publishing may be explained by their underrepresentation in STEM as researchers. If the percentage of women authors in submitted articles is higher than in published articles, this would support a bias in peer review explanation, indicating that policies and practices at publishing firms would need to be addressed to support inclusion.

Most studies to date have analyzed gender gaps in STEM authorship across various disciplines globally, revealing that women are significantly underrepresented in fields like physics, computer science, and engineering, while the life sciences and social sciences show smaller, though still present, disparities (West et al., 2013; Holman et al., 2018; Huang et al., 2020). However, these analyses have largely focused on international data sources or major publishers like Elsevier (2017) and there is a notable lack of studies specifically examining gender disparities within journals published by Canadian publishers. Our study builds on this research and was guided by research questions that have been explored in previous research but have not yet been examined exclusively using data from Canadian Science Publishing journals. The present study provides the first large-scale analysis of gender representation in Canadian STEM journals. Using specialized software, we estimated the gender of authors of scientific articles and analyzed women's and men's representation in submitted and published articles, as solo and co-authors, and as first and last authors using over 67,000 articles from 24 Canadian Science Publishing journals. This study offers new insights into the extent to which Canada has achieved gender equity in scientific publishing—or whether persistent disparities reflect ongoing challenges of imbalanced representation in STEM.

## Method

The not-for-profit organization Canadian Science Publishing (CSP) is Canada's largest publisher of international science journals (https://cdnsciencepub.com/about), and at the time of data collection, consisted of 24 different journals across 7 different disciplines (Table 1). As of 2024, two journals, Anthropocene Coasts and Geomatica, while included in our study, are no longer associated with CSP, and the Journal of Unmanned Vehicle Systems has been renamed to Drone Systems and Applications.

**Table 1.** Canadian Science Publishing journals and numbers of manuscripts by discipline

| Discipline | Journal Name | Number of Manuscripts Analyzed | Years Analyzed |
|---|---|---|---|
| Environmental Sciences | Environmental Reviews* | 40 | 2011-2013 |
| | Anthropocene Coasts | 51 | 2017-2021 |
| Biological Sciences | Canadian Journal of Fisheries and Aquatic Sciences | 4855 | 2011-2021 |
| | Canadian Journal of Forest Research | 4392 | 2011-2021 |
| | Botany | 2171 | 2011-2021 |
| | Canadian Journal of Zoology | 2632 | 2011-2021 |
| Agricultural Sciences | Canadian Journal of Plant Science | 2824 | 2011-2021 |
| | Canadian Journal of Animal Science | 1597 | 2011-2021 |
| | Canadian Journal of Soil Science | 1067 | 2011-2021 |
| Engineering | Canadian Geotechnical Journal | 559 | 2011-2021 |
| | Transactions of the Canadian Society for Mechanical Engineering (TCSME) | 820 | 2017-2021 |
| | Canadian Journal of Civil Engineering | 5515 | 2011-2021 |
| Biomedical sciences | Biochemistry and Cell Biology | 2298 | 2011-2021 |
| | Canadian Journal of Microbiology | 6456 | 2011-2021 |
| | Genome | 1781 | 2011-2021 |
| | Canadian Journal of Physiology and Pharmacology | 5525 | 2011-2021 |
| | Applied Physiology, Nutrition, and Metabolism | 6127 | 2011-2021 |
| Physical Sciences | Canadian Journal of Chemistry | 5058 | 2011-2021 |
| | Canadian Journal of Earth Sciences | 1887 | 2011-2021 |
| | Canadian Journal of Physics | 6533 | 2011-2021 |
| | Geomatica | 50 | 2018-2021 |
| Multidisciplinary | FACETS | 269 | 2015-2021 |
| | Drone Systems and Applications (formerly Journal of Unmanned Vehicle Systems) | 173 | 2013-2020 |
| | Arctic Science | 188 | 2015-2020 |

*Notes*. The data from Environmental Reviews was limited to 40 valid articles from 2011 to 2013, due to its focus on review-type submissions, which were outside the primary scope of our analysis (research articles).
Time ranges vary by journal due to different data availability.

Bibliometric information including author names from all 24 journals listed was shared by CSP to researchers for analysis. Data included submissions from approximately September 2010 to February 2021 (each journal had slightly different date ranges, see Table 1). The original dataset contained 80,371 entries, which was reduced to 67,601 entries after data cleaning and applying exclusion criteria described below. This study used secondary data from CSP, analyzed in its original form without direct interaction with human subjects. The study protocol was reviewed and approved by the Wilfrid Laurier University Research Ethics Board (approval number 9285).

**Data Cleaning**

Entries were excluded if they lacked data on editorial decisions to publish or not, and only articles with final decisions between 2011-2021 were retained. Only entries classified as research articles were analyzed. All entries not classified as research articles (e.g., reviews, notes, topical communications, invited reviews, technical notes, systematic reviews, rapid communications, symposium, and data papers, etc.), duplicate entries (e.g., multiple instances of the same manuscript ID or repeated names associated with the same article), and cases where there was insufficient information to determine gender (e.g., where only initials were provided) were removed prior to analysis. The remaining entries constituted our valid dataset of all submitted papers (including those that were accepted for publication and those that were rejected; $N$ = 67, 601). We performed calculations using this dataset, as well as the dataset consisting solely of accepted, published articles ($N$ = 18,755).

**Gender Name Inference**

While gender is a complex social construct that extends beyond a binary framework (e.g., Coburn et al., 2024; Hyde et al., 2018; Nussbaumer et al., 2025; Risman et al., 2022; Saguy et al., 2021; Vijlbrief et al., 2020; Ward & Lucas, 2024), most bibliometric research utilizes a 'man' and 'woman' binary notion of gender identity due to data limitations (e.g., Jaramillo et al., 2025; Larivière et al., 2013; West et al., 2013). As in previous studies, we assigned gender based on names using standard software, acknowledging that this approach has inherent constraints, such as reinforcing a narrow view of gender, and errors in assigning gender accurately. Nonetheless, this analysis provides a critical first step in understanding gender disparities in Canadian STEM publishing and serves as a foundation for future research that explores gender representation in scientific publication.

To assess gender representation among authors, we utilized NamSor API version 2. (https://namsor.com), a name recognition software that assigns gender to authors based on names. The dataset provided by CSP did not include any gender information, necessitating this step. NamSor employs machine learning to estimate the gender associated with names, covering all languages, alphabets, countries, and regions. NamSor was also selected due to its ability to handle a large volume of names. To assign gender, the software returns a score between -1 and 1, where a score of -1 indicates 'man' with 100% certainty, and a score of 1 represents 'woman' with 100% certainty. A score in between -1 and 0 indicates 'man,' with decreasing certainty as the number approaches 0, and a score between 0 and 1 indicates 'woman' with increasing certainty as the number approaches 1. Names that are commonly used for both men and women receive a lower certainty score than other names.

NamSor also produces individual accuracy values for each name that it analyzes, represented as a percentage. For example, an 80% accuracy value can be interpreted such that the program is 80% sure that the gender assigned to the name is correct. NamSor's accuracy in gender assignment has been validated through comparisons with other gender identification software, demonstrating a reasonable degree of precision and global coverage. This software has implemented rigorous protocols to assess its performance, showing high recall (i.e., few unknowns) and high accuracy (i.e., few false positives) across various countries, including the United States, Canada, Mexico, Russia, Japan, China, and several European nations (Mattauch et al., 2020; Santamaria & Mihaljević, 2018; Sebo, 2023). In our study, we evaluated NamSor's certainty for each journal included in our analysis by computing the average certainty and standard deviation based on individual certainty values for each name. The standard deviation represents the spread of the certainty scores from the average certainty score. The average certainty over all journals was 86%, and the average standard deviation over all journals was 16% (Table 2). This accuracy testing provides a foundation for our analysis of gender representation in Canadian Science Publishing journals.

**Table 2.** Average NamSor certainty and associated standard deviation

| **Journal Name** | **Average Certainty** | **Standard Deviation** |
|---|---|---|
| Anthropocene Coasts | 78.1% | 18.5% |
| Applied Physiology, Nutrition, and Metabolism | 92.9% | 13.2% |
| Arctic Science | 95.1% | 10.2% |
| Biochemistry and Cell Biology | 78.2% | 18.7% |
| Botany | 88.8% | 16.2% |
| Canadian Geotechnical Journal | 80.0% | 18.8% |
| Canadian Journal of Animal Science | 82.5% | 18.4% |
| Canadian Journal of Civil Engineering | 86.6% | 16.9% |
| Canadian Journal of Chemistry | 84.5% | 17.9% |
| Canadian Journal of Earth Sciences | 85.0% | 17.8% |

| Canadian Journal of Fisheries and Aquatic Sciences | 91.9% | 14.0% |
|---|---|---|
| Canadian Journal of Forest Research | 88.2% | 16.4% |
| Canadian Journal of Microbiology | 82.1% | 18.4% |
| Canadian Journal of Physics | 81.9% | 18.9% |
| Canadian Journal of Physiology and Pharmacology | 82.9% | 18.3% |
| Canadian Journal of Plant Science | 81.7% | 18.4% |
| Canadian Journal of Soil Science | 84.0% | 17.8% |
| Canadian Journal of Zoology | 92.0% | 14.2% |
| Environmental Reviews | 93.9% | 11.3% |
| FACETS | 93.1% | 12.7% |
| Genome | 78.2% | 18.8% |
| Geomatica | 84.9% | 17.2% |
| Drone Systems and Applications (formerly Journal of Unmanned Vehicle Systems) | 92.1% | 14.0% |
| Transactions of the Canadian Society for Mechanical Engineering (TCSME) | 75.8% | 18.6% |

Acceptance rates and average authors per published article were calculated. For gender-related data, we determined the percentage of women first authors, along with women submitting, corresponding, and last authors. We examined team composition, categorizing manuscripts into those written by solo women, solo men, multiple men, multiple women, and mixed-gender groups. This process was repeated for both all submissions and the subset of published articles. Author team composition was assessed at the manuscript level (i.e., solo women, solo men, co-authored by women, co-authored by men, and co-authored by women and men). Separately, proportions of women authors were calculated at the author level, defined as the total number of women authors divided by the total number of all authors across manuscripts. We investigated three key questions to uncover patterns in how men and women were represented across Canadian Science Publishing journals:

1. How does gender representation differ across Canadian Science Publishing journals and academic disciplines?

2. How does the composition of authorship teams influence publication outcomes for women in Canadian Science Publishing journals?

3.How are women represented in different authorship roles in Canadian Science Publishing journals?

## Results & Discussion

Our key findings illustrated that men authors were more prevalent across all CSP journals compared to women authors, and while women were not abundantly published, there was no difference between the percentage of women submitting and publishing their science. Across disciplines, our findings indicated that women authors were more prevalent in biomedical sciences, and less so in engineering, environmental, physical, and agricultural sciences. In other words, these analyses do not necessarily provide direct evidence of systematic gender bias in the review process. Although some research has assessed potential gender bias in the academic review process, results are somewhat mixed such that gender bias against women authors in review is sometimes evident, and sometimes not (Ceci et al., 2023; Fox et al., 2023; Kern-Goldberger et al., 2022; Squazzoni et al., 2021; Tregenza et al., 2008). Our findings suggest that the lower representation of women in the scientific research ecosystem and as submitting authors better explains the gender publishing gap than gender bias in review alone. However, this was not an experimental study, and we cannot rule out bias completely. Publishers can monitor representation, policies, and practices to ensure that potential biases are mitigated. Efforts to increase women's representation as researchers (e.g., guarding against bias in hiring, implementing gender-inclusive workplace policies; Moss-Racusin et al., 2021) and accessibility of publishing (e.g., mentorship centered editorial models, open-access platforms; Wongvorachan, 2025) are necessary to reduce the gender publishing gap – which is arguably more challenging than implementing policy and practices to remove bias from the publishing process.

Consistent with previous research, our findings highlight that science publishing is collaborative; across all CSP journals (using the mean percentages across years and journals), most publications were co-authored by teams of men and women. Findings also showed that the percentage of publications co-authored by men was more than six times larger than the percentage of publications co-authored by women. When women do publish alone in CSP journals, they do so more in certain journals (i.e., the Canadian Journal of Physics). However, when grouping by discipline, we see that in the physical sciences, the percentage of solo men authors was three times the size of the percentage of solo women authors. In some CSP journals (e.g., Environmental Reviews), there were no solo woman-authored publications at all. However, it should be noted that the number of manuscripts analyzed varied between Environmental Reviews and other journals.

The present analysis showed patterns consistent with past research such that women were less represented than men in first and last authorship positions. In just 4 of 24 CSP journals, the percentage of women first authors reached or exceeded 50%. Our analysis also revealed stark disciplinary differences in women as first authors; in biomedical sciences, there were approximately equal percentages of men and women first authors, however, in disciplines where women are less represented such as engineering, this percentage fell by more than half.

**How does gender representation differ across Canadian Science Publishing journals and academic disciplines?** The representation of women varied widely across disciplines, but notably, not a single journal had a larger percentage of women authors than men. Looking at the mean across all disciplines and journals in our analysis, women represented 34% of authors who submitted manuscripts and 33% of authors who published manuscripts (Figure 1[1]). Men comprised 66% of authors who submitted manuscripts, and 67% of authors who published manuscripts (Figures S1 and S2, see supplementary materials for additional figures). Compared to global estimates indicating that women comprise approximately 41% of active researchers — or 33% in the physical sciences (Elsevier, 2024) — our findings showed that women were even less prevalent authors in Canadian Science Publishing journals. However, our analysis also revealed no difference between the percentage of women who submitted and who were published authors in CSP journals, which is encouraging and particularly noteworthy.

[1] For all figures, the numbers next to the bars represent the exact percentages.

In terms of disciplinary differences, our analysis showed that biomedical sciences had the highest representation of women authors (44%) among published manuscripts (Figure 2), which is close to global trends wherein women represent ~48% of authors in related fields (i.e., medicine, biochemistry, genetics, and molecular biology; Van der Linden et al., 2024). Conversely, we observed that women were underrepresented as published authors in environmental sciences (27%) and engineering (21%), while globally, data show that women constitute 42% of published authors in environmental sciences and 28% of published authors in engineering (Van der Linden et al., 2024). Engineering is a prime example where women's low representation in Canada (i.e., about 22%; Engineers Canada, 2021) and around the globe (20%; World Economic Forum, 2021) mirrors the low percentage of published women authors in engineering (21% in the current study).

Among Canadian Science Publishing journals, we also observed marked gender disparities between men and women in the physical and agricultural sciences. Women comprised just 26% of authors in physical sciences, 34% in agricultural sciences, 35% in multidisciplinary sciences publications, and 34% in biological sciences. In comparison to some previous research, these values are lower than global averages (i.e., ranging from 28% of publications in physical sciences to 44% of publications in immunology and microbiology sciences; Van der Linden et al., 2024), and are consistent with data illustrating that women are less published in sciences in which they are also underrepresented in general.

**Figure 1.** Percentage of submitted and published manuscripts with women authors by journal.

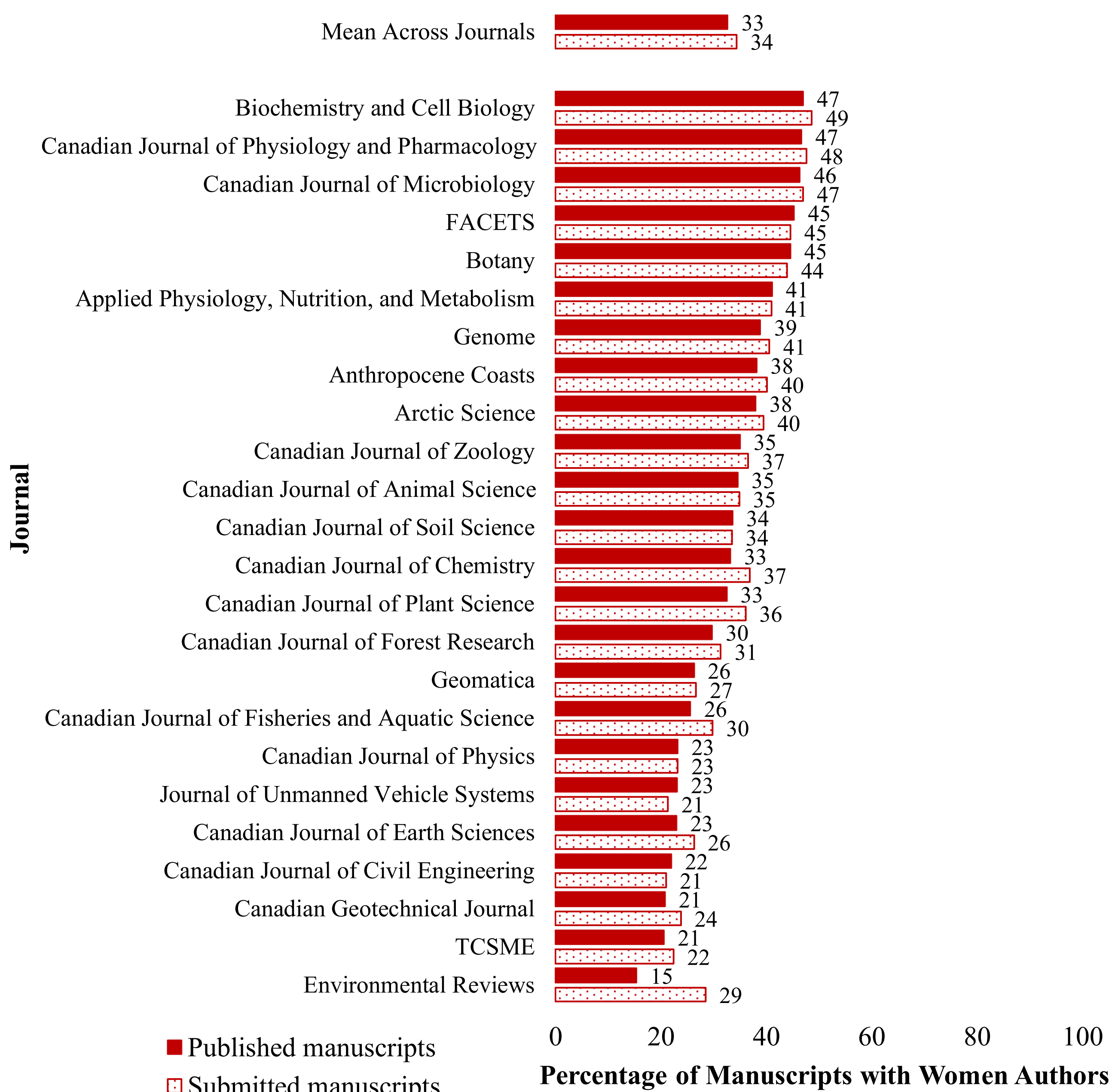

**Figure 2.** Percentage of submitted and published manuscripts with women authors by discipline.

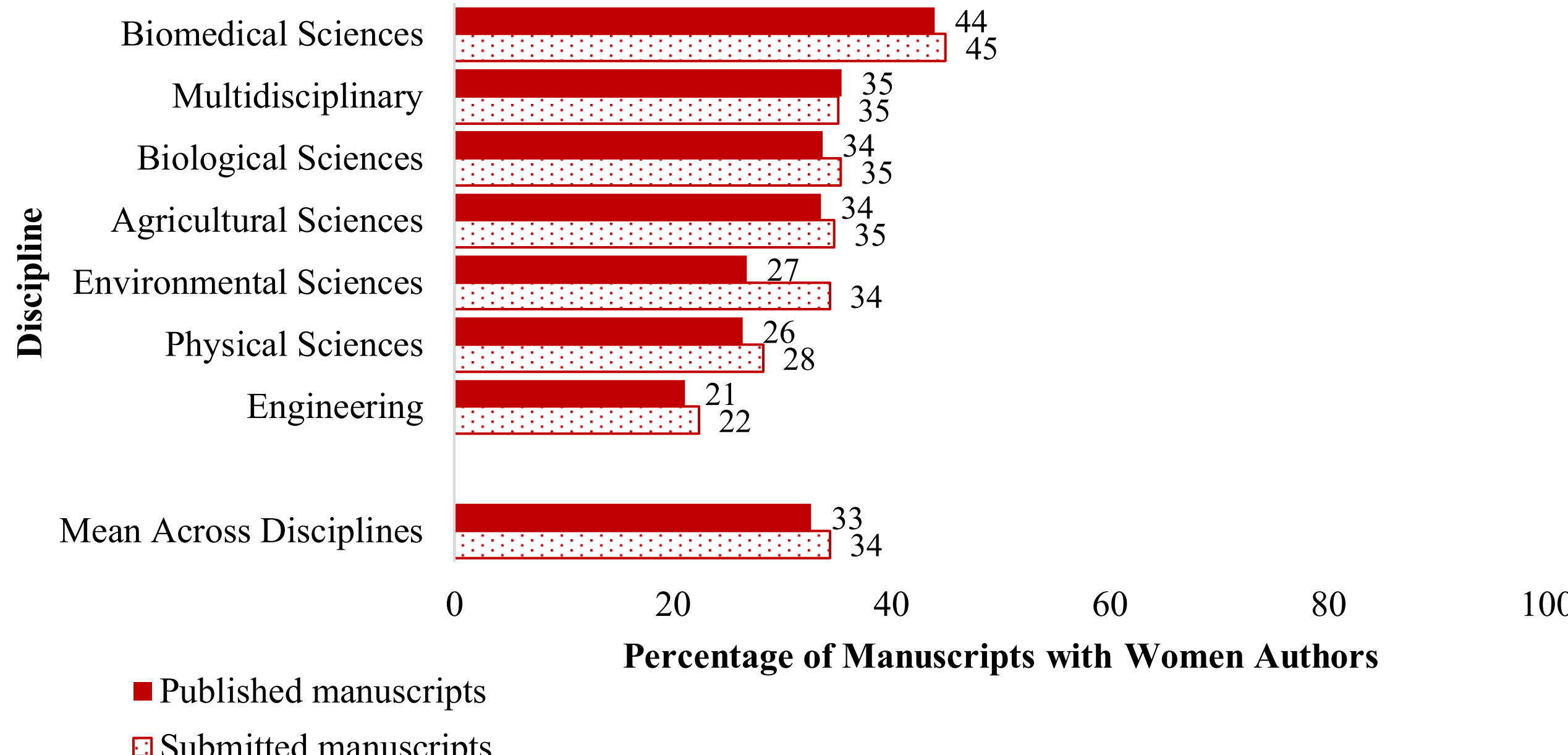

**How does the composition of authorship teams influence publication outcomes for women in Canadian Science Publishing journals and across academic disciplines?**

We examined the gender composition of authors in Canadian Science Publishing journals, assessing percentages of solo authors and co-authored manuscripts. Of all published manuscripts, co-authored manuscripts by men and women were the most common (63%), followed by those co-authored by men only (26%), solo-authored by men (6%), co-authored by women, (4%) and lastly, solo-authored by women (2%: Figures 3-6). Higher publication rates for co-authored (men + women) publications were found in the biomedical sciences (74%), agricultural sciences (73%), multidisciplinary sciences, and biological sciences (66%; Figure 5). This pattern of higher publication rates for co-authored articles is consistent with trends in science, whereby collaboration is related to increased research output, greater citational visibility, and more impact (Baker et al., 2019, Shen et. al, 2021).

Manuscripts co-authored exclusively by women accounted for an average of 4% of publications overall, varying across disciplines from 7% in biomedical sciences to only 2% in engineering (Figure 3). Manuscripts authored by women alone comprised just 2% of publications overall (Figure 4). While the low percentage of solo-woman authors is consistent with trends found in previous research (Kwiek & Roszka, 2022), in the present study we do not have the necessary data to identify causal factors contributing to the gender gap in scientific authorship. However, previous research paints a complex picture of how a multitude of factors such as caregiving, having a partner in academia, gender stereotypes, limited access to collaborative networks, and citational practices (e.g., self-citation), influence women's high-impact authorship (Dworkin et al., 2020; Cech & Blair-Loy, 2010; Kwiek & Roszka, 2022; Eagly et al., 2019; Uhly et al, 2015).

The patterns of authorship for women in co-authored manuscripts across disciplines generally aligned with the representation of women in these fields, showing a higher percentage of publications in journals within the biomedical sciences, multidisciplinary sciences, and biological sciences (Figure 5). In contrast, lower percentages of co-authored women publications were observed in engineering and physical sciences, a pattern that aligns with the underrepresentation of women within these fields (Figure 5). This pattern shifts when examining solo-authored manuscripts by women, where higher percentages of publications were found in the physical sciences, particularly in journals like the Canadian Journal of Physics (Figure 7). This deviation suggests that in certain fields, such as physics, individual women researchers have managed to carve out niches, enabling them to produce solo publications. Indeed, the comparison of solo-woman authors with the broader trends in women's authorship revealed that in fields such as physics, solo-woman authorship was notably higher than the overall trend of women's authorship (Figure 7). In the Canadian Journal of Physics, solo-woman authors accounted for 5% of all publications, a percentage that more than doubles the 2% of solo-woman authors observed across all journals in the dataset (Figures 3-6).

We found variability in percentage of articles solo-authored by women. When compared to the overall representation of women authors in the dataset, solo-woman authors were more represented in journals like Canadian Journal of Physics, Geomatica, Biochemistry and Cell Biology, and Canadian Journal of Physiology and Pharmacology (Figure 7). In contrast, journals such as Anthropocene Coasts, Canadian Geotechnical Journal, Environmental Reviews, and Canadian Journal of Microbiology showed substantially lower percentages of solo-woman authors. Some of these journals, like Anthropocene Coasts and Environmental Reviews, had no solo-woman authors at all (Figure 7). Kwiek and Roszka (2022) demonstrated that while women scientists across disciplines published solo less often than men, when they did publish alone, they did so with greater frequency, across man-dominated and woman-dominated disciplines. This finding aligns with ours, wherein solo women in the physical sciences (especially physics) appeared to produce more solo publications, suggesting that in certain disciplines, solo authorship by women may be pronounced despite the issue of widespread underrepresentation.

**Figure 3.** Percent of published manuscripts co-authored by women and men, co-authored by women only, or co-authored by men only, by journal.

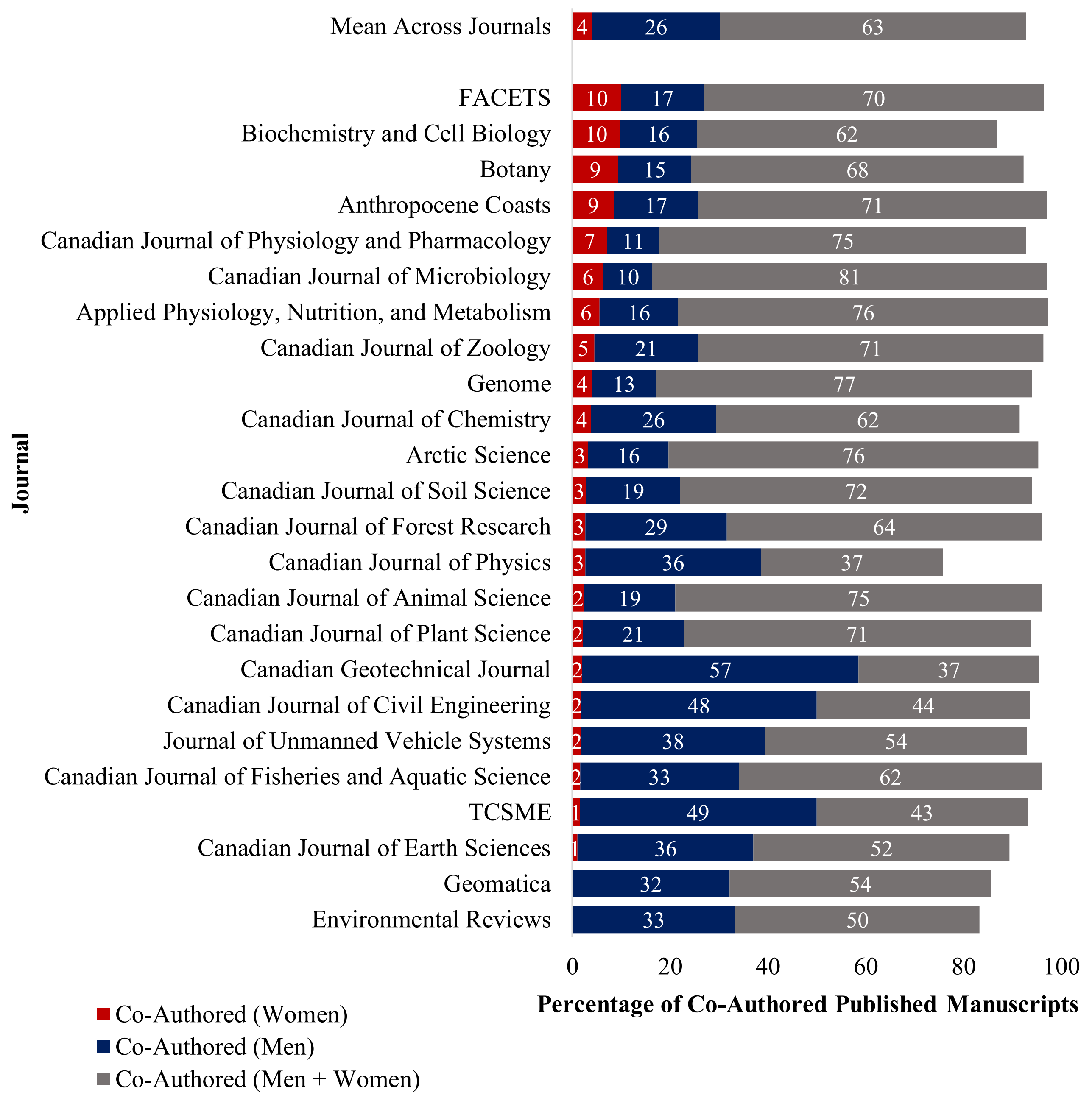

**Fig. 3. Note.** This data only represents co-authored works, excluding solo-authored publications.

**Figure 4.** Percentage of published manuscripts with solo authors by gender and journal.

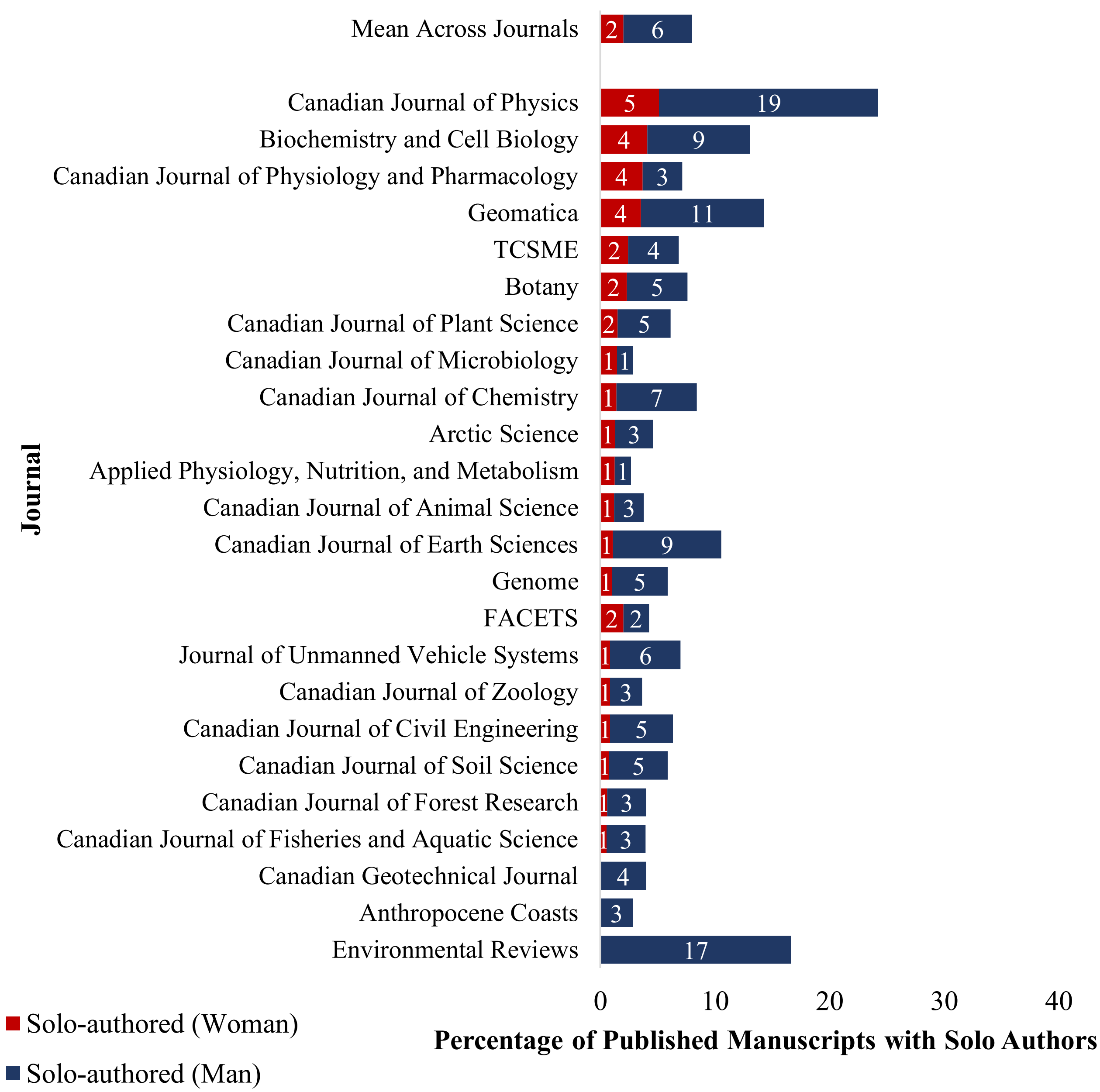

**Fig. 4. Note.** The x axis has been modified for ease of interpretation.

**Figure 5.** Percentage of manuscripts that were co-authored by women and men, co-authored by women only, or co-authored by men only, by discipline.

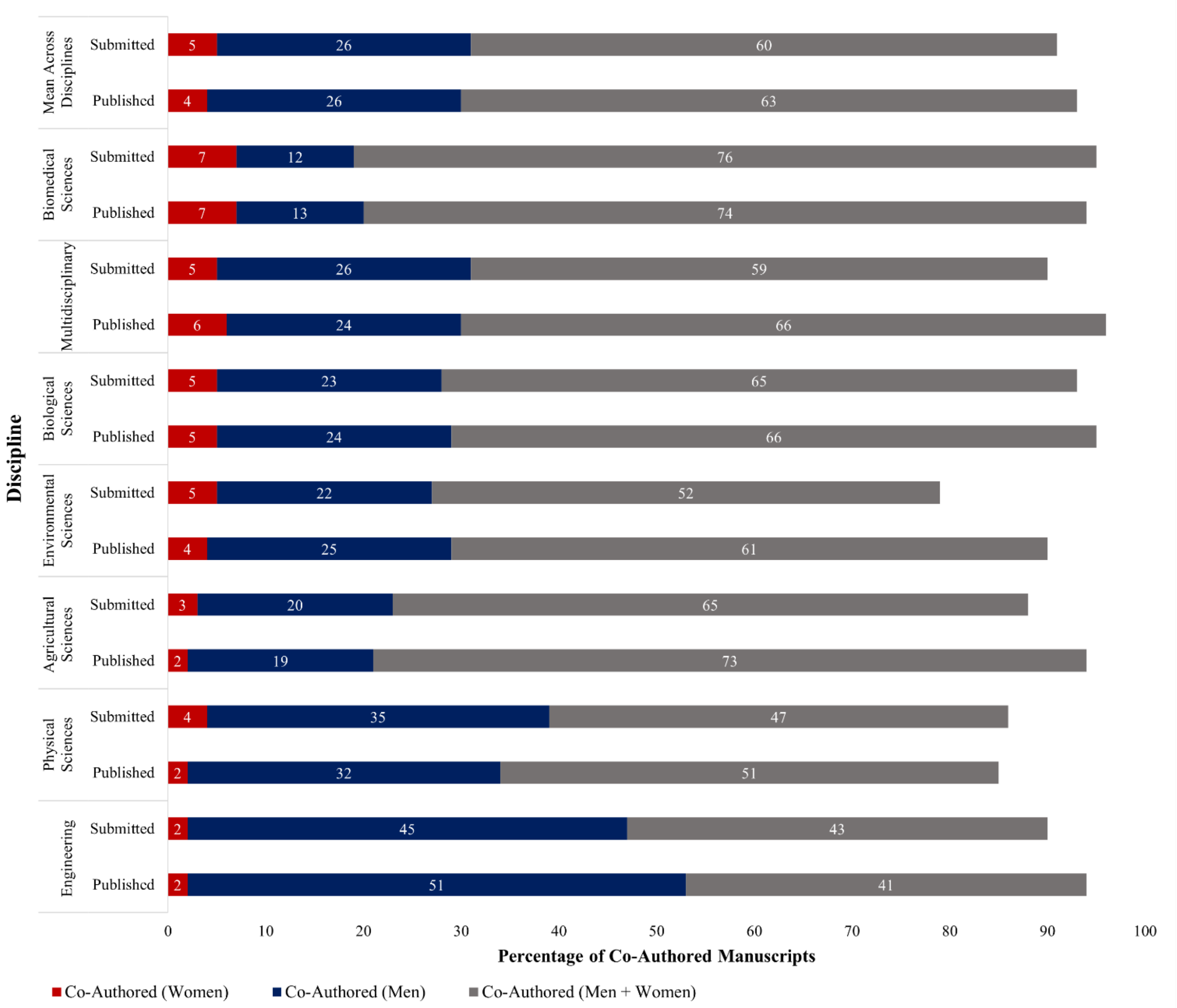

**Figure 6.** Percentage of solo-authored manuscripts by gender and discipline.

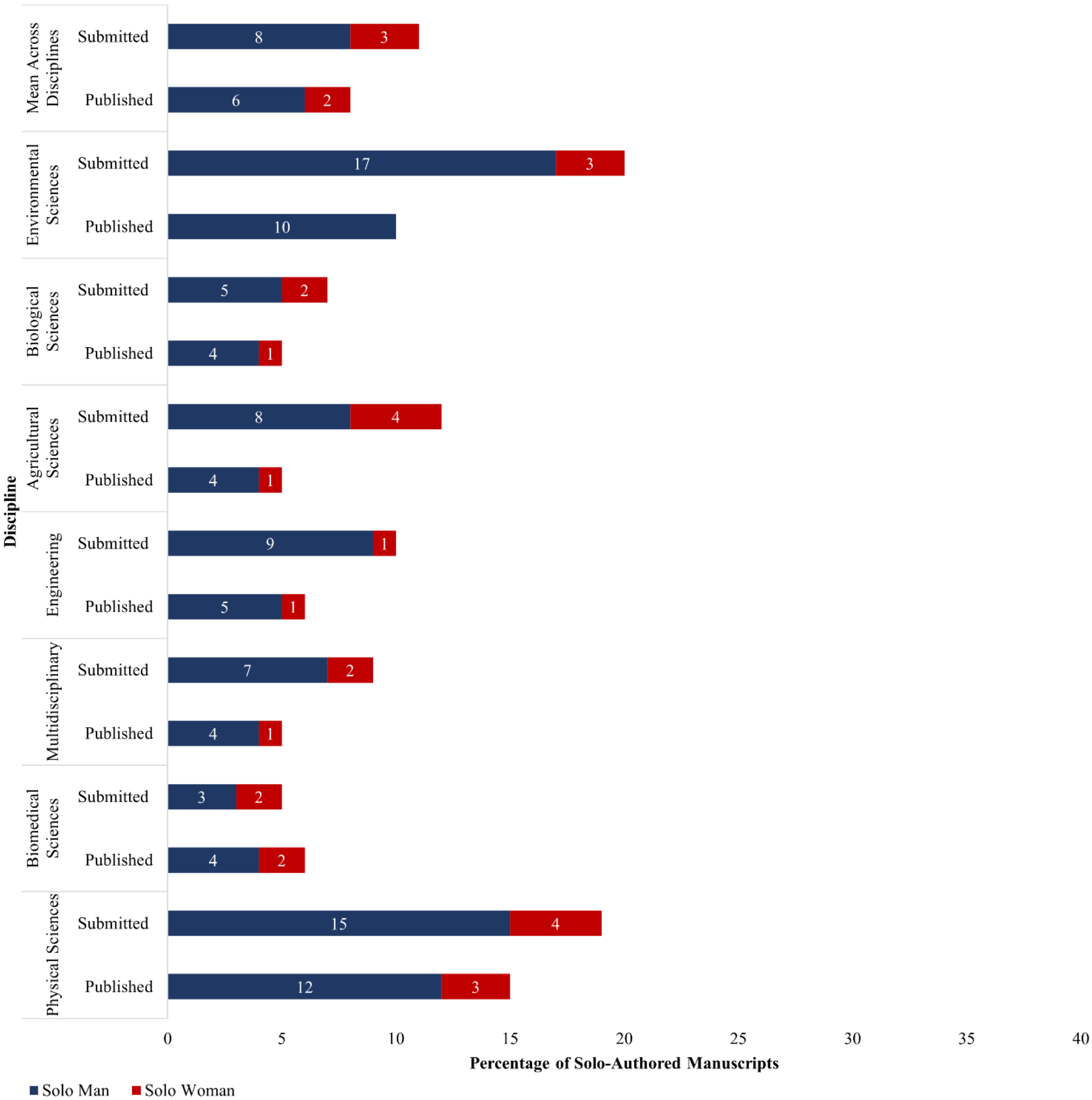

**Fig. 6. Note.** The x axis has been modified for ease of interpretation.

**Figure 7.** Percentage of published women authors and publications solo-authored by women by journal.

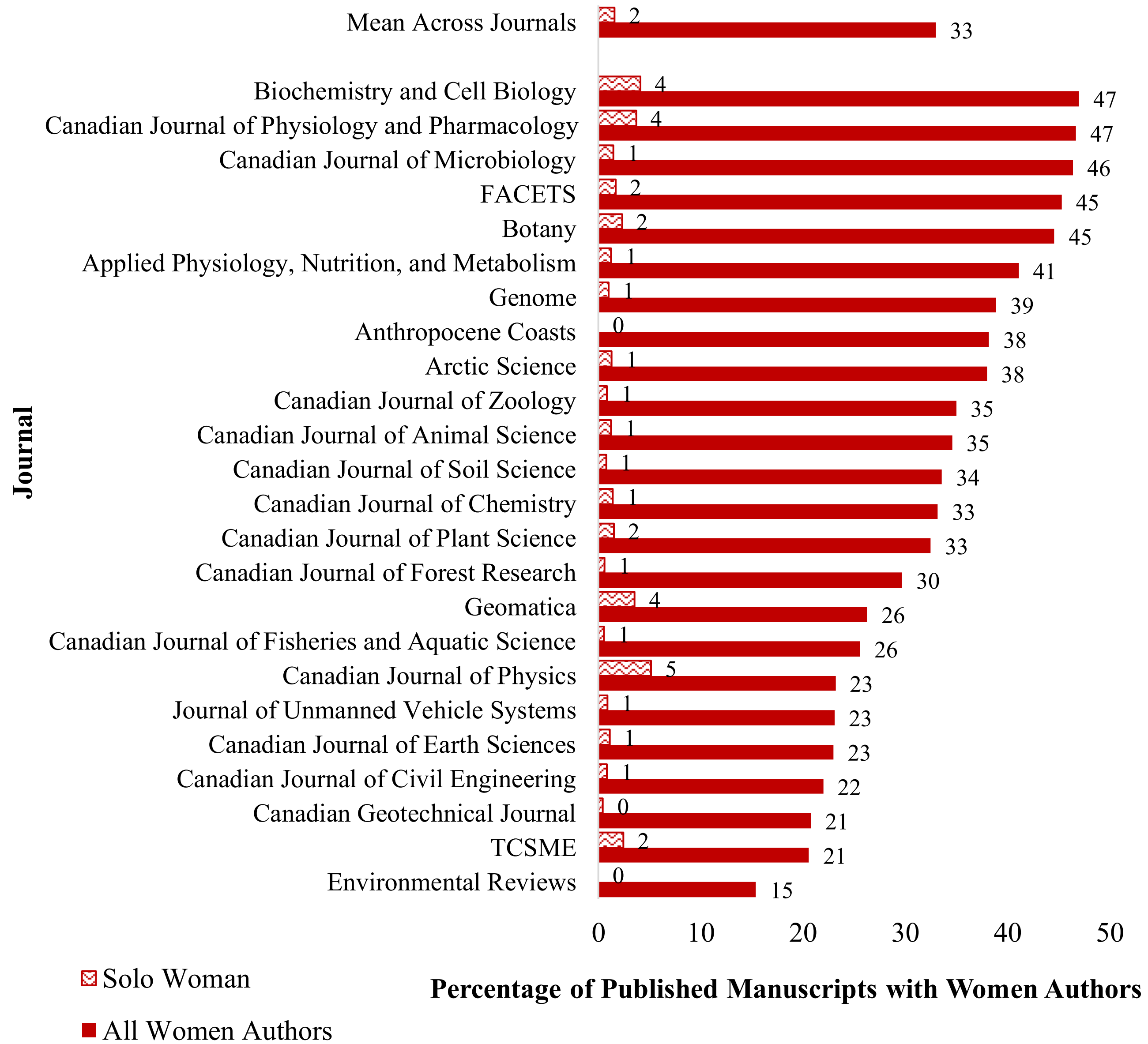

**Fig.7. Note.** The 'all women authors' category includes both solo and co-authored publications. The x axis has been modified for ease of interpretation.

**Figure 8.** Percentage of published men authors and publications solo-authored by men by journal.

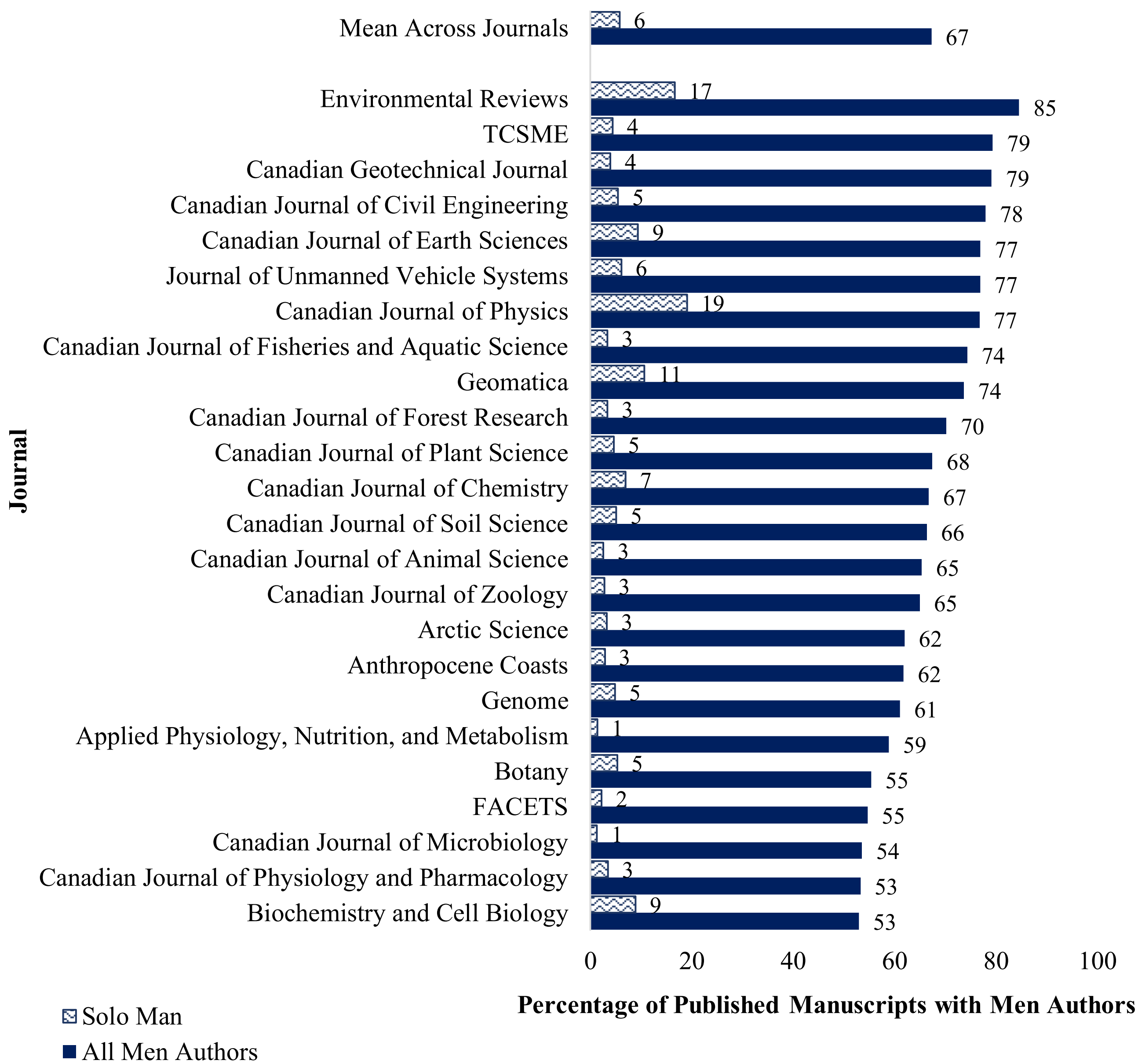

**Fig. 8. Note.** The ‘all men authors’ category includes both solo and co-authored publications.

**How are women represented in different authorship roles in Canadian Science Publishing journals and across academic disciplines?**

Recalling that first and last author positions are generally important in scientific publications, we analysed the percentages of women and men in first and last author positions across journals. Across all journals, women were 36% of first authors, compared to 64% that were men (Figures 9 and 10). This parallels the gender representation in overall authorship (Figures 1 and 2). The representation of women in other consequential authorship positions showed a similar pattern, whereby women comprised 30% of submitting authors, 31% of corresponding authors, and 27% of last authors (see Figures S3-S5 in supplementary materials). In several journals, women comprised 50% or more of first authors (i.e., FACETS, Canadian Journal of Microbiology, Canadian Journal of Physiology and Pharmacology, and Biochemistry of Cell Biology). By discipline, we found a higher percentage of women first authors in fields in which women are more represented: in biomedical sciences (48%), multidisciplinary sciences (41%), and biological sciences (41%). We also observed fewer women first authors in disciplines wherein women are less represented: in engineering journals (22%) and environmental science journals (17%). Our findings are therefore consistent with international trends demonstrating that women first and last authors are less common than men first and last authors (e.g., Hallas, 2025; Kwiek & Roszka, 2022). Given how critical first and last authorship positions are to academic success, it will be important for future research to disentangle the extent to which authorship order reflects who is represented in science versus potential barriers in the publication process.

**Figure 9.** Percentage of man and woman first-authors by gender and journal.

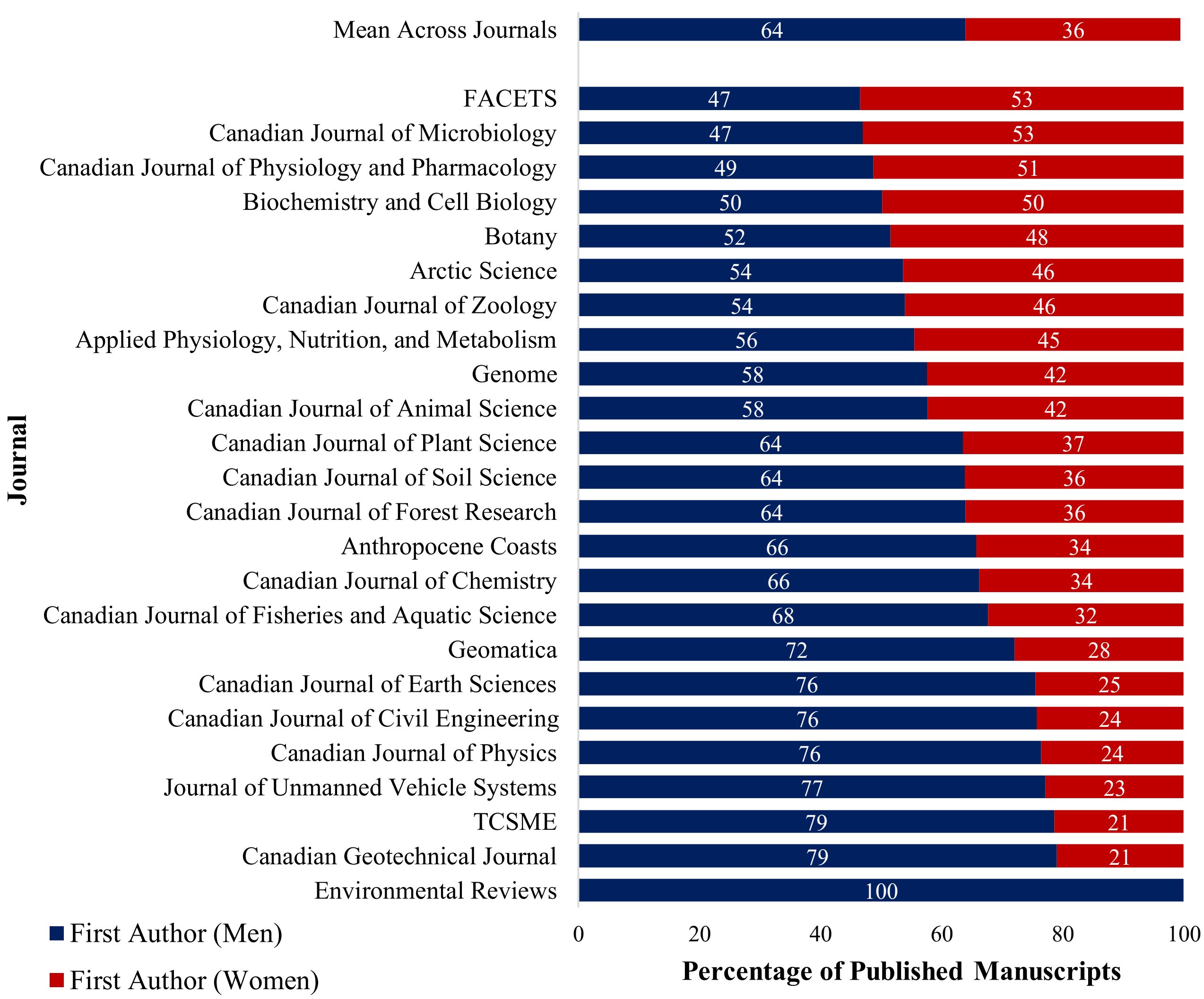

**Figure 10.** Percentage of men and women first-authors by gender and discipline.

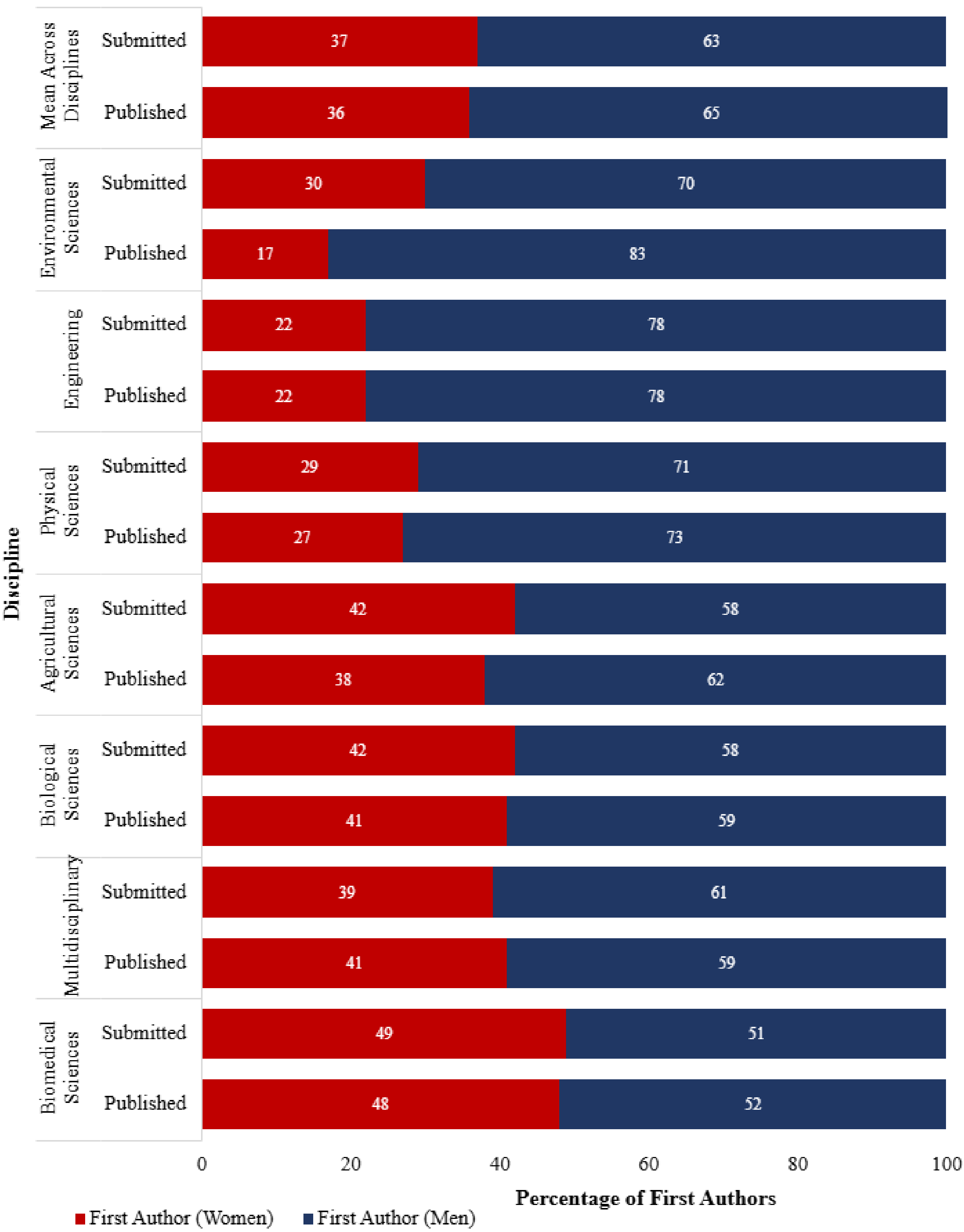

## Conclusion

The Canadian government, and academic and research institutions, have invested significant time and funding into inclusion in science (Government of Canada 2025a, 2025b; NSERC 2025a, 2025b). As such, all components of the Canadian research ecosystem should be assessed and monitored regarding demographic representation and bias. Yet research is lacking when it comes to understanding women's representation in the broader Canadian science publishing landscape. This study offers new insights into gender representation within journals published by Canadian Science Publishing, the country's largest scientific publisher. In general, our findings are consistent with previous work showing women authors are less represented in scientific publications than men (Elsevier, 2017; Elsevier, 2024; Huang et al., 2020), and, perhaps unsurprisingly, this disparity was most notable in fields in which women were less represented (i.e., physical sciences, engineering, mathematics). While it is reasonable that the pool of available scientists is reflected in bibliometric data, this parallel only emphasizes the need to increase gender representation in science writ large while simultaneously improving access to inclusive publication practices. Because our analysis did not find a difference between the percentages of women submitting and being published, this suggests that the underrepresentation of women in the overall pool of authors is more likely driving the gender gap, rather than bias in specific aspects of the Canadian Science Publishing journal review processes. However, this was not an experimental study, so we cannot rule out bias altogether.

Analysing this unique data set, we found that women overall published less than men in CSP journals, but this varied substantially between journal and by discipline. Our analysis also reiterates that science is a team sport; most publications were co-authored by teams of men and women, and co-authoring was particularly prevalent in biomedical sciences, multidisciplinary sciences, and biological sciences. This finding supports existing literature that suggests gender-diverse teams enhance scientific productivity and contribute to a more inclusive research environment (Bear & Woolley, 2011; Casad et al., 2021; Ley & Hamilton, 2008; Nielsen et al., 2017). On the other hand, it is crucial to consider the potential implications of solo authorship for women in science. While team collaboration may be productive, publishing as a solo woman in a man-dominated STEM field may be one way to propel women's success, given solo authorship makes it clear about who completed scientific work (Sarsons et al., 2021). While women solo authors were the least represented in the dataset, their representation differed across journals and disciplines, highlighting the varied nature of women's participation and publication experiences across scientific fields. The present analysis supports previous research showing that women were underrepresented as last authors—positions that are significant because they typically indicate leadership and senior responsibility in a publication (e.g., Hallas, 2025). As noted in past research, the lower representation of women in influential authorship positions limits their visibility and opportunities for advancement within the academic community (Shen, 2013, 2018; West et al., 2013).

Our study had several limitations. First, we relied on an imperfect automated gender classification software based on names (NamSor). While NamSor is quite effective at assigning gender to Western names, especially Latin or Anglophone names, it is often less accurate when determining the gender of authors with African, Arabic, or Asian names (Mattauch et al., 2020, Santamaria & Mihaljević, 2018, Sebo, 2023). Thus, it will be important for future research to conduct comparable analyses using different software to replicate results. Second, the current study utilized a 'man' and 'woman' gender binary categorization, which is inherently restrictive to gender non-conforming people and reinforces a cisnormative conceptualization of gender. Given some research in the Canadian context suggests that gender and sexual minorities may be more represented in some STEM disciplines than the general Canadian population (e.g., in physics; Hennessey & Smolina et al., 2025), it will be critical to build systems that allow a more accurate categorization of gender, preferably through a self-reporting mechanism. Third, our analysis was limited to publications in Canadian Science Publishing journals from 2011 to 2021, which restricts our ability to assess long-term trends, which will be essential to understand progress. Finally, we did not conduct intersectional analyses of author identity, as many journals do not yet collect information on gender identity, race, sexuality, disability, or class required to do so. To allow for intersectional bibliometric analyses, it will be necessary for journals to collect standardized intersectional demographic data over time.

Longer-term data on women authors in Canadian Science Publishing journals—the country's largest scientific publisher—would allow continuous monitoring of women's representation as authors, and track progress (or lack thereof) over time. Data collected post-COVID-19 will provide insight into the impacts of the pandemic on publishing for women and men. Given Canada's commitment to EDI in STEM, we hope that the representation of women and other underrepresented groups will increase. To ensure that the STEM environment is equitable and inclusive for all researchers, a commitment to continuous monitoring of potential publication bias is crucial. As the STEM workforce diversifies, we must ensure that the pool of submitting and published authors reflects that diversity. Future research can also investigate the factors that may have contributed to our results. That is, what practices and policies are in place at Canadian Science Publishing journals that are effective at mitigating publication bias? Further, what practices and policies may increase submissions from women and other groups who are underrepresented in STEM? Although beyond the scope of the current dataset, other work has begun to examine the inclusive practices that help encourage submissions and remove bias from publishing (e.g., providing guidance on reducing bias to reviewers, implementing inclusive hiring best practices when hiring editors, using inclusive language in communications with authors and reviewers; See Jack et al., 2023).

Our research suggests women's underrepresentation in STEM as an explanation for their underrepresentation as authors. Gender gaps in science cannot be traced to a single cause but instead result from a complex interplay of social, cultural, and systemic factors—ranging from individual experiences, such as bias, to structural issues, like the absence of policies promoting gender equity in publishing and lack of gender inclusive policies in STEM institutions (see Avolio et al., 2020 and Schmader, 2023 for reviews). All members of the scientific community have a role to play in increasing representation. Publishers can support this endeavor by continuing to monitor outcomes and examining their practices to ensure that the Canadian publishing ecosystem is free from bias and supports researchers in publishing their innovative and excellent work.

**Statement of Author Contributions:**

1. Eden J. Hennessey led the writing and revising of the submitted manuscript, provided supervision on analysis, data visualization, and editing.
2. Amanda Desnoyers conducted the primary analyses included in the submitted manuscript, did data analysis, made charts, formal data analysis, data visualization, and editing.
3. Margie Christ cleaned and prepared all author data for analysis, conducted preliminary analyses, and assisted with initial manuscript writing and data visualization.
4. Adrianna Tassone prepared data, verified analyses, and edited the manuscript.
5. Skye Hennessey prepared data, verified analyses, and provided project management and research team coordination.
6. Bianca Dreyer conducted data classification and initial calculations.
7. Alex Jay conducted data classification and initial calculations.
8. Patricia Sanchez conducted data classification and initial calculations.
9. Shohini Ghose conceptualized the project, initiated the partnership with Canadian Science Publishing, and provided supervision on data analysis and manuscript editing.

**Acknowledgments**

We are grateful to Canadian Science Publishing, who provided the data that made this work possible. We gratefully acknowledge the numerous authors of scientific publications, without whose work this analysis would not have been possible. The WinS team acknowledges that their work is done on the shared traditional territory of the Neutral, Anishnaabe and Haudenosaunee peoples. This land is part of the Dish with One Spoon Treaty between the Haudenosaunee and Anishnaabe peoples and symbolizes the agreement to share, protect our resources and not to engage in conflict. This project was supported by the Natural Sciences and Engineering Research Council of Canada's Chairs for Women in Science and Engineering grant (WIS2U 606168-2025).

**Data Availability**

The data referenced in this publication are the property of Canadian Science Publishing and were used under strict confidentiality agreements. Due to the presence of personally identifiable information, the dataset cannot be shared publicly or with third parties.

**Competing Interests**

The authors declare there are no competing interests.

**Inclusion and Ethics Statement**

As a research team, we worked together across nations and institutions as a demographically diverse group (i.e., diversity in gender, race, sexual orientation, and disability as well as discipline), in collaboration with the Canadian Science Publishing (CSP). We cite national and international literature relevant to the publishing context through meta-analytic and specific reports as well as research relevant to underrepresented communities (i.e., women in science). The study was reviewed and approved by the Wilfrid Laurier Research Ethics Board (REB# 9285).

## Supplementary Information

### Additional Figures

**Figure S1.** Percentage of submitted and published manuscripts with men authors by journal.

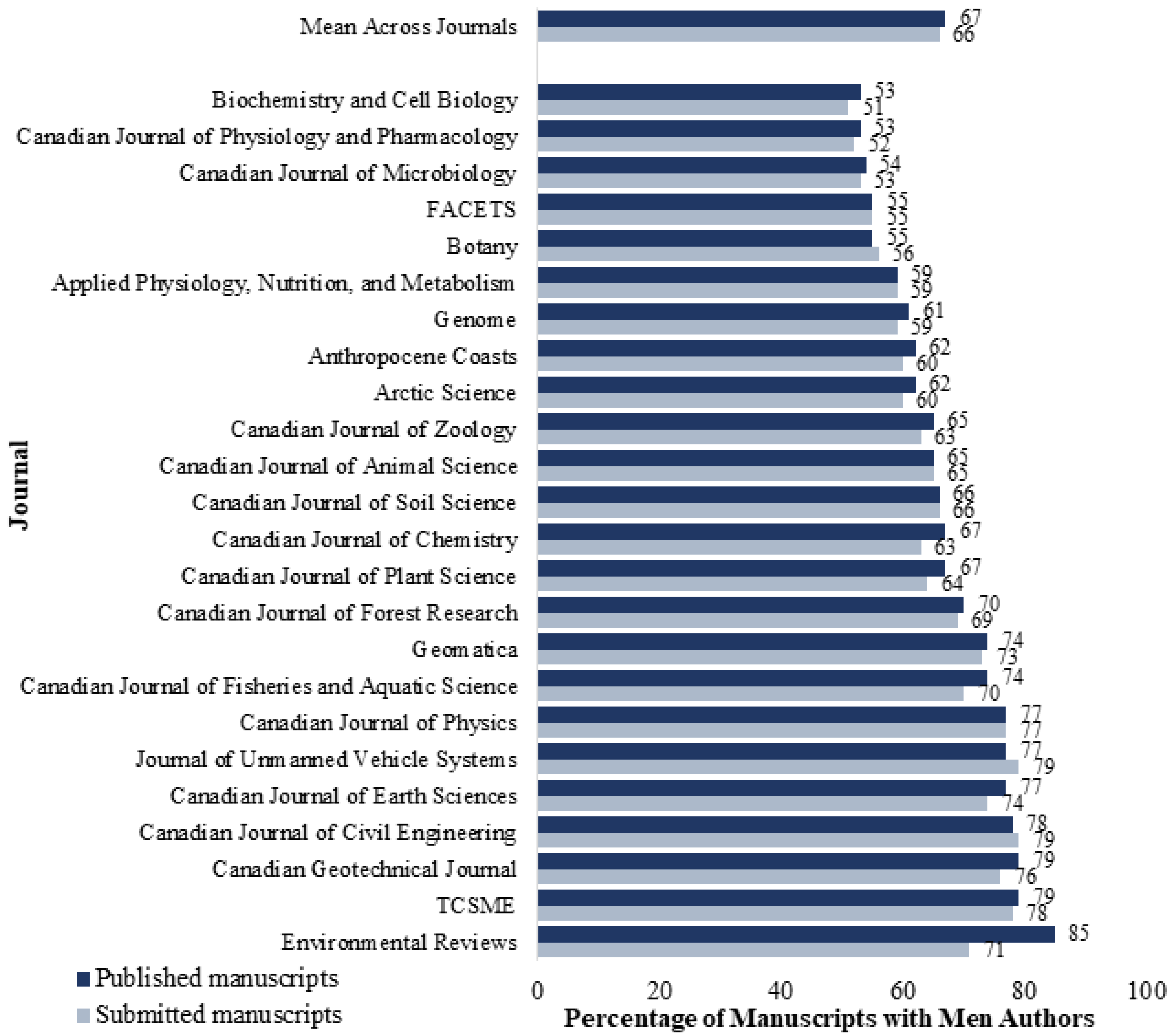

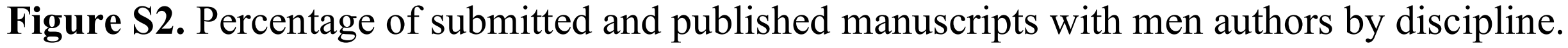

**Figure S2.** Percentage of submitted and published manuscripts with men authors by discipline.

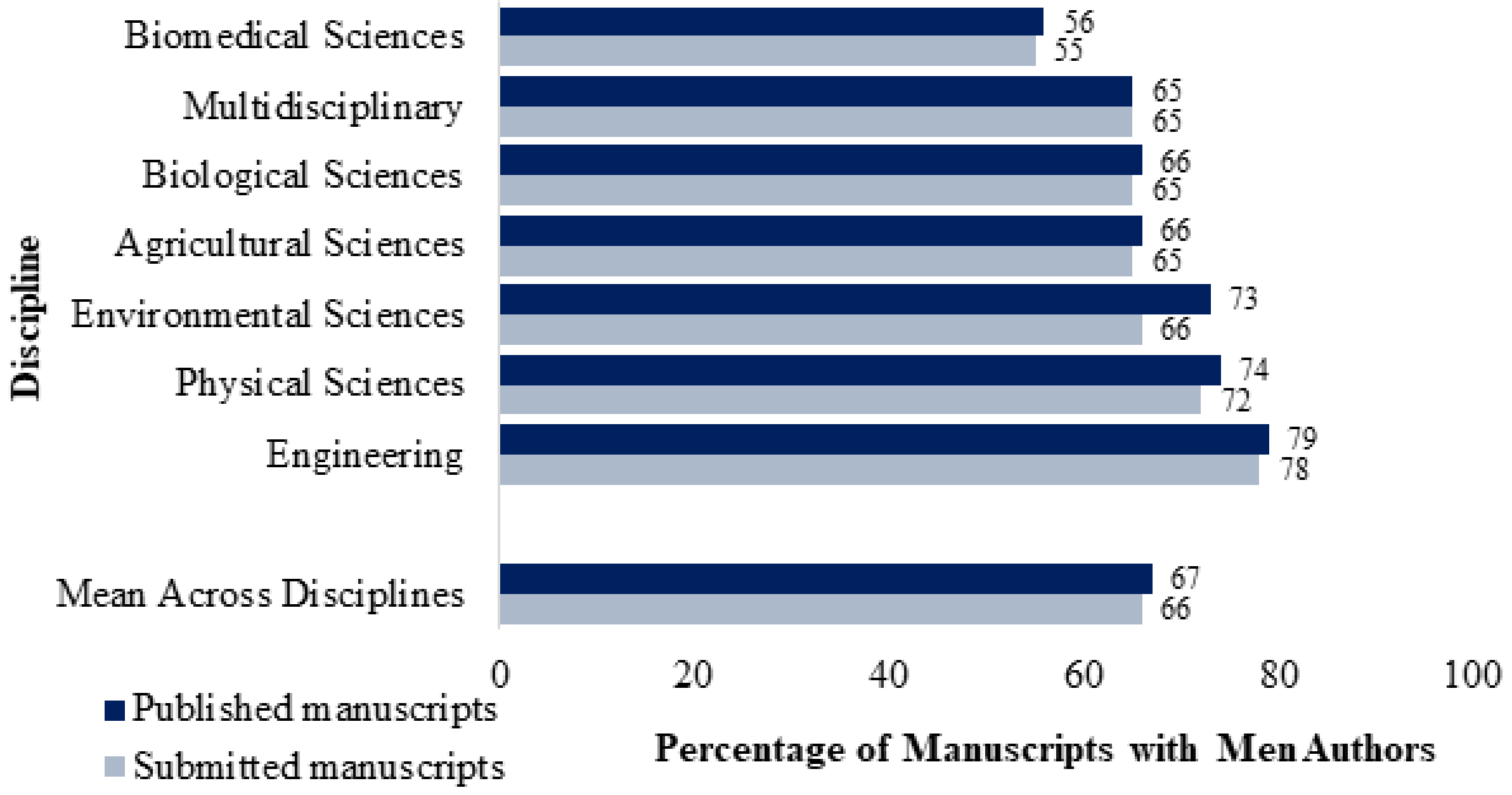

**Figure S3.** Percent of manuscripts with a woman submitting author by discipline.

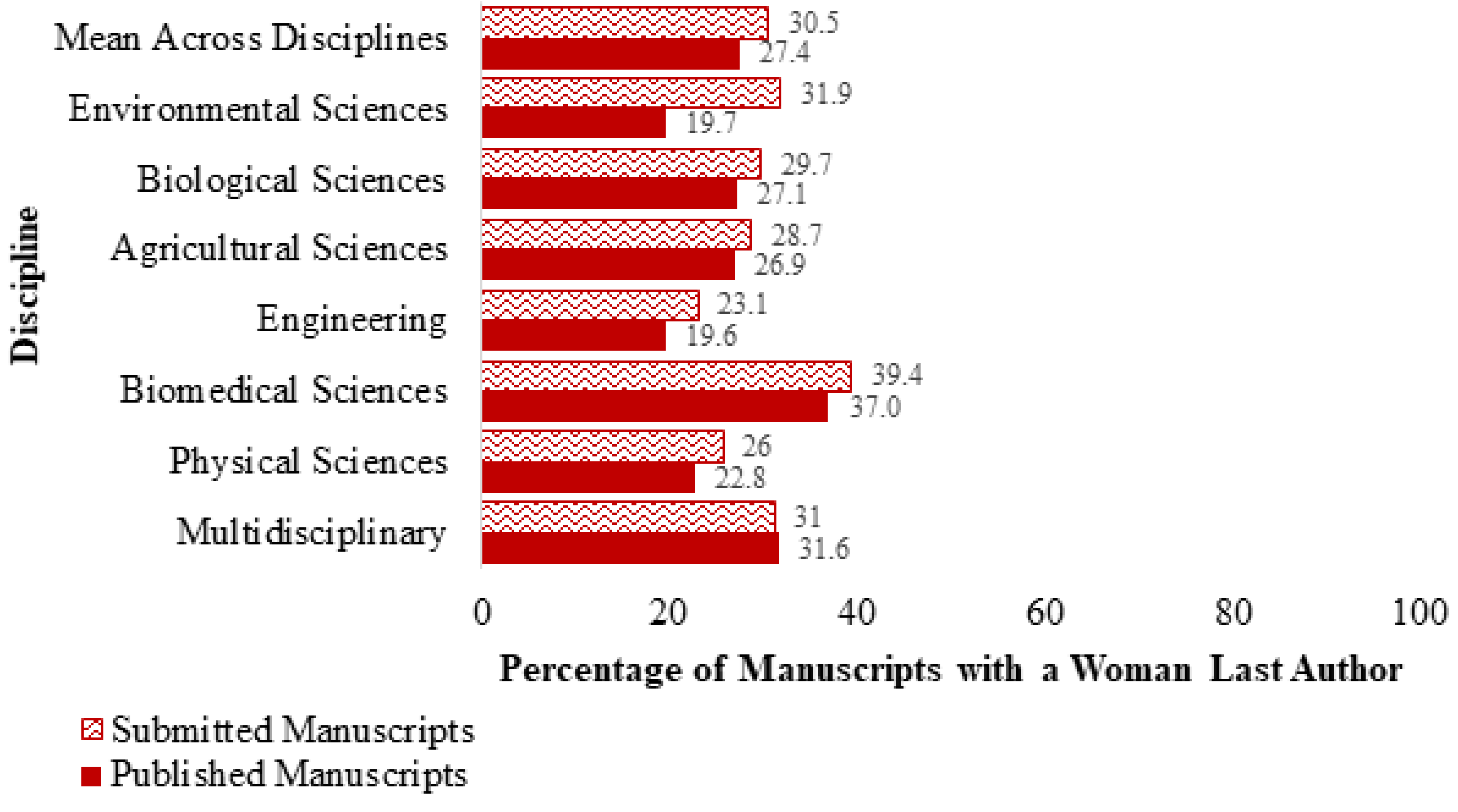

**Figure S4.** Percent of manuscripts with a woman corresponding author by discipline.

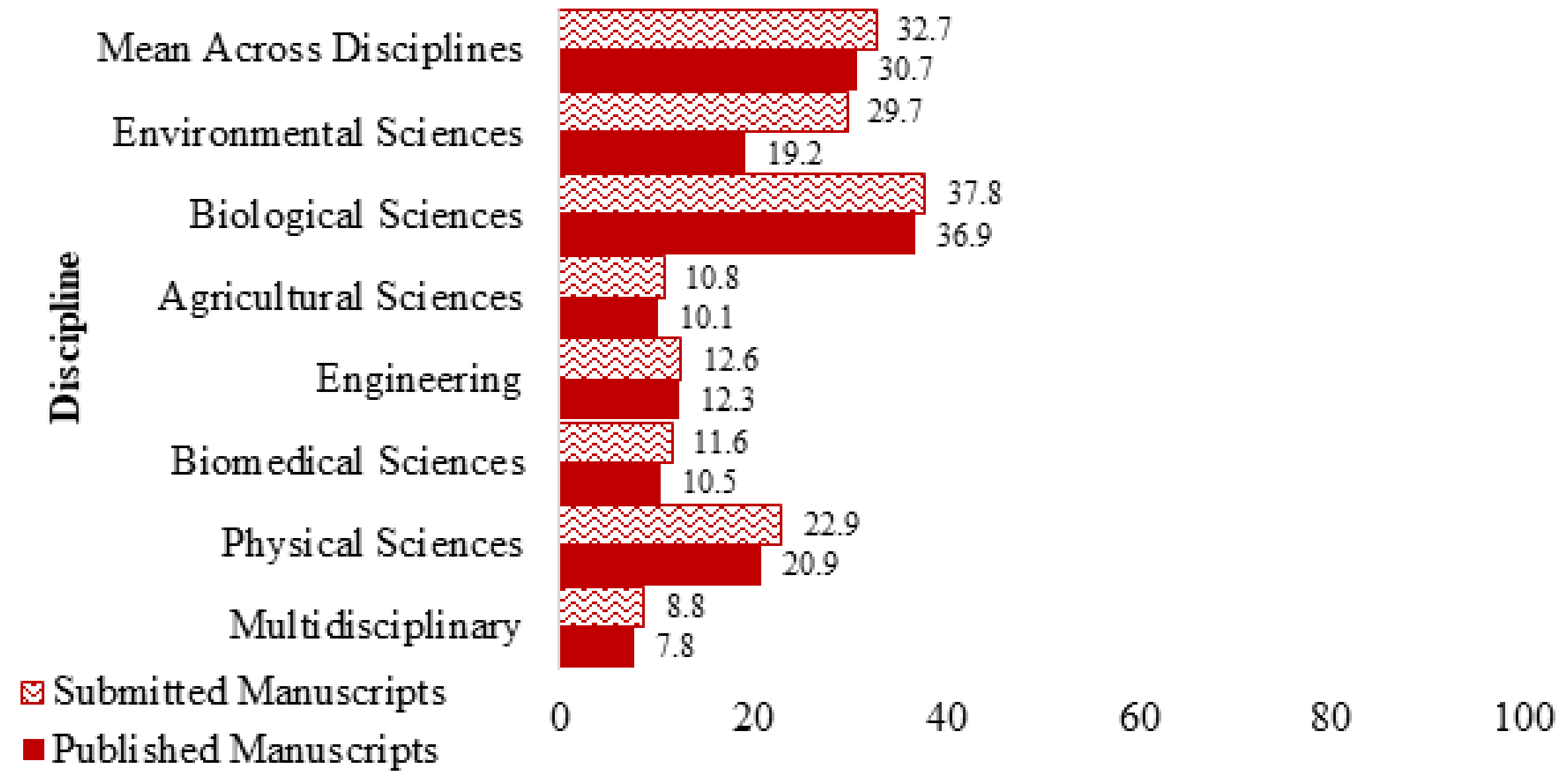

**Figure S5.** Percent of manuscripts with a woman last author by discipline.

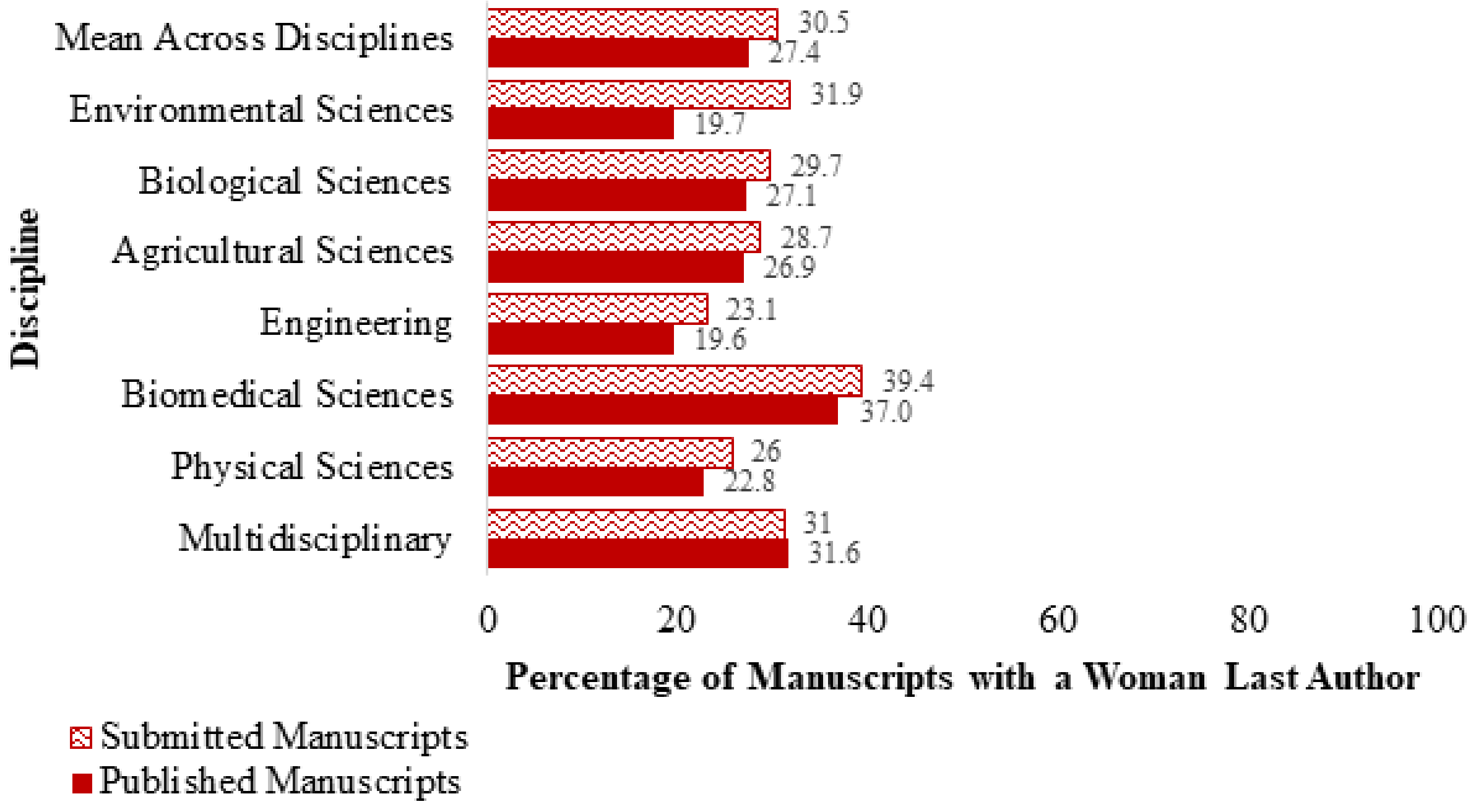